\documentclass[review]{elsarticle}
\usepackage[top=2.75cm,bottom=2.75cm,left=2.75cm,right=2.75cm]{geometry}
\usepackage{natbib}
\usepackage{amsmath} 
\usepackage{amsfonts}
\usepackage{graphicx}
\biboptions{numbers,sort&compress}
\begin{document}
	\begin{frontmatter}
\title{Semi-numerical solution for a fractal telegraphic dual-porosity fluid flow model}
\author[mymainaddress]{E. C. Herrera-Hern\'andez\corref{mycorrespondingauthor}}
\cortext[mycorrespondingauthor]{Corresponding author}
\ead{erik.herrera@cidesi.edu.mx}
\author[mysecondaryaddress]{C. G. Aguilar-Madera}
\author[mymainaddress]{D. P. Luis}
\author[mythirdaddress]{D. Hern\'andez} 
\author[myfourthaddress]{R. G. Camacho-Vel\'azquez}
\address[mymainaddress]{CONACYT-Centro de Ingenier\'ia y Desarrollo Industrial, Av. Playa pie de la cuesta 702, Desarrollo Sn. Pablo, 76125, Quer\'etaro, Qro., Mexico}
\address[mysecondaryaddress]{Facultad de Ciencias de la Tierra, Universidad Aut\'onoma de Nuevo Le\'on, Ex-hacienda de Guadalupe, 67700, Linares, N.L., Mexico}
\address[mythirdaddress]{Maestr\'ia en Ciencias de la Complejidad, Universidad Aut\'onoma de la Ciudad de M\'exico, Ciudad de M\'exico, Laboratorio Nacional de Ciencias de la Complejidad, Ciudad de M\'exico}
\address[myfourthaddress]{Corporate Direction of Technology Research and Development, PEMEX}

\begin{abstract}
In this work, we present a semi-numerical solution of a fractal telegraphic dual-porosity fluid flow model. It combines Laplace transform and finite difference schemes. The Laplace transform handles the time variable whereas the finite difference method deals with the spatial coordinate. This semi-numerical scheme is not restricted by space discretization and allows the computation of a solution at any time without compromising numerical stability or the mass conservation principle. Our formulation results in a non-analytically-solvable second-order differential equation whose numerical treatment outcomes in a tri-diagonal linear algebraic system. Moreover, we describe comparisons between semi-numerical and semi-analytical solutions for particular cases. Results agree well with those from semi-analytic solutions. Furthermore, we expose a parametric analysis from the coupled model in order to show the effects of relevant parameters on pressure profiles and flow rates for the case where neither analytic nor semi-analytic solutions are available.
\end{abstract}
\begin{keyword}
	\texttt{semi-numerical solution\sep fractal dual-porosity\sep fluid flow model\sep numerical Laplace inversion\sep finite differences method\sep Laplace transform}
\end{keyword}
\end{frontmatter}
\section{Introduction}
The Laplace transform (LT) method has been widely used to solve problems in several areas of science and engineering. After proper transformation of the space variable, the application of LT to dynamic models is useful for finding semi-analytic solutions for many real dynamic problems in the so-called {\it Laplace domain} \cite{EH49,WR63,B88,CY93,PCK00,CFV08,MPBI13}. Those solutions need to be inverted in order to obtain appropriate solutions in the time domain; if analytical inversion is not possible, then numerical procedures such as the Stehfest \cite{S70} or de Hoog \cite{HKS82} algorithms are employed. Sometimes, just the short- or long-term system behavior is of interest, and for this purpose asymptotic approaches are commonly used to avoid the inversion needed for complete solutions \cite{FC01,FC03}. For solute transport and fluid flow problems in porous media, such as aquifers \cite{B88}, oil, and geothermal reservoirs, the LT method helps to find semi-analytic solutions \cite{EH49,CFV08,WR63,RRT14}, exhibiting many advantages over purely numerical procedures. The main example is where subsequent inverse modeling will be implemented in order to retrieve model parameters of interest. However, for more complex problems, the semi-analytic solution is often not easy to obtain or does not exist, and other alternatives need to be explored. In this context, numerical solutions are commonly used; but if they are not properly handled, some restrictions related to the spatial discretization process appear. Specifically, these frequently include numerical dispersion, instabilities due to the discretization procedure, and violations of the mass conservation principle.

Recently, hybrid methods combining LT and finite differences (LTFD) have successfully been applied to solve problems of slightly compressible fluid flow in oil reservoirs \cite{MR91,MMH94} two-phase fluid flow problems \cite{HO14}, and even for the solution of advection-dispersion-type models in the economy \cite{TM13} or in heat conduction problems\cite{CC88}. This sort of hybrid method presents some valuable characteristics which are absent in numerical solutions: i) there are no restrictions on the time variable as it is approximated in the Laplace domain, so the solution can be generated for any specific value, and ii) the initial conditions and other time-dependent properties are easily treated in the Laplace domain. However, it is worth mentioning that the LTFD method has some disadvantages compared to the purely-numerical approach: i) as time approaches zero, the Laplace parameter tends to diverge and oscillating solutions could appear, and ii) the LTFD method is not robust when solutions behave in a non-smooth way \cite{WZ15}. As the aim of this work is to present the model solution for non-Fickian and fractal double-porosity/single-permeability reservoirs, and to analyze the LTFD methodology as a potential tool for modeling fluid reservoirs, we put special attention on the details  described above.

In this work, through the combination of the Laplace transform and finite difference methods, we developed and validated semi-numerical solutions for fractal telegraphic double-porosity models previously derived in \cite{HNV13}. In the cited work, only solutions for particular cases were reported; they were used to validate our model solutions. We solved the complete model, whose solution, to the best of our knowledge, has not been reported before in the literature. Further, a parametric analysis of physical parameters was carried out. The paper is outlined as follows: in Sec. \ref{model}, we present the mathematical model along with initial and boundary conditions. Sec. \ref{method} presents a detailed description of the LTFD method. Results and discussions are shown in Sec. \ref{resdis}. Finally, concluding remarks and perspectives are summarized in Sec. \ref{conc}.
\section{Theory}\label{model}
\subsection{The double-porosity model}
As in the pioneering work of Warren and Root \cite{WR63}, the physical model consists of a porous medium constituted by two media: one where fluid flows and the other where fluid is stored. The former is associated with the connected fracture network and the latter to the rock matrix. One of the main differences between the Warren and Root model and the one solved in this paper is that here the porous medium is assumed to be a percolating cluster, where backbone structure plays the role of the fracture network and dead-ends play the role of the rock matrix. Fluid motion occurs in the backbone network whereas dead-end structures can participate or not in the flow, depending on physical parameters. Thus, the dimensionless mathematical model describing the fluid behavior in isotropic fractal reservoirs due to participation of the fracture network and rock matrix is given by \cite{HNV13}

\begin{eqnarray}
(1-\omega)\frac{\partial h_{1}}{\partial t} = \frac{\lambda}{r^{d_{de}-d}}\left( h_{2}-h_{1}\right)\label{e1}\\
\begin{split}
\omega\tau\frac{\partial^{2} h_{2}}{\partial t^{2}}+\left(\omega+\lambda\tau r^{d-d_{bb}}\right)\frac{\partial h_{2}}{\partial t} = \frac{1}{r^{d_{bb}-1}}\frac{\partial}{\partial r}\left(r^{\beta}\frac{\partial h_{2}}{\partial r}\right)+\\
\left(\lambda\tau r^{d-d_{bb}}-\left[1-\omega\right]r^{d_{de}-d_{bb}}\right)\frac{\partial h_{1}}{\partial t}\label{e2},\\
 \quad \textrm{where}\quad \beta=d_{bb}-1-\theta
\end{split}
\end{eqnarray}
In the last equation, $h_{1}=h_{1}(r,t)$ is the dimensionless dead-ends  hydraulic head and $h_{2}=h_{2}(r,t)$ is the dimensionless hydraulic head inside the percolation backbone. The parameters $d_{bb}$ and $d_{de}$ are the fractal dimensions of the backbone network and dead-end structure, respectively, $d$ is the Euclidean dimension in which the system is embedded, $\omega$ is the backbone storativity ratio, and $\tau$ is the relaxation time associated with fluid flow within the backbone fracture network. The parameter $\lambda$ quantifies the fluid exchanged between the percolation backbone and the dead-end structure while $\theta$ is related to the connectivity degree of the backbone network. More details of the model can be found in \cite{HNV13}.

 Eq. (\ref{e1}) represents the hydraulic head evolution inside the dead-ends, including mass exchange between dead-end and fracture networks. Meanwhile, Eq. (\ref{e2}) is related to head evolution inside the backbone due to gradient-driven flux and dead-end contributions. Note from Eq. (\ref{e1}) that fluid accumulation inside dead-ends depends only on the fluid gradient between the backbone and the dead-ends. Furthermore, the amount of fluid passing from the dead-ends to the backbone is determined by the exchange parameter $\lambda$ and position within the porous medium. In this sense, the discharge point is not connected to dead-ends, but rather is only connected to the backbone structure. On the other hand, fluid inside backbones flows through porous media at a finite velocity, and its motion depends on the flux-driven term and the mass exchange with the dead-ends, as described in Eq. (\ref{e2}).

\subsection{Initial and boundary conditions}
As the model is of second-order in time and space for the backbone equation and of first-order in time for  the dead-end equation, we need two initial conditions and two boundary conditions for Eq. (\ref{e2}) and one initial condition for Eq. (\ref{e1}). Writing the initial conditions as general space-dependent functions, we have:
\begin{equation}\label{e3}
 h_{1}(t=0,r)  = g_{1}(r), \quad 1\le r\le L,
\end{equation}

\begin{equation}\label{e4}
h_{2}(t=0,r) = g_{2}(r)\quad \textrm{and},\quad \frac{\partial h_{2}(t=0,r)}{\partial t} = g_{3}(r)
\end{equation}

The boundary conditions depend on the operating conditions being modeled. In this work, we define two cases: i) a bounded reservoir with constant head at the wellbore and ii) a bounded reservoir with asymptotically constant flow rate at the wellbore. For the first case, we set constant head at the inner boundary as follows:

\begin{equation}\label{e5}
h(r=1,t) = h_{w},
\end{equation}
which is a Dirichlet-type condition, and for the outer boundary we impose zero flux (a Neumann-type boundary condition), i.e.,
\begin{equation}\label{e6}
\frac{\partial h}{\partial r}\big{|}_{r=L} =0.
\end{equation}
The instantaneous flow rate is evaluated considering the memory of the flow caused by the non-locality in time as follows:
\begin{equation}\label{e7}
Q(t) = -\frac{1}{\tau}\int_{0}^{t}\exp\left(-\frac{t-\hat{t}}{\tau}\right)\,\frac{\partial h}{\partial r}{\big{|}}_{r=1}\,d\hat{t}.
\end{equation}
Note that, to evaluate the amount of fluid discharged at any time, it is necessary to integrate the solution from the beginning. Eq. (\ref{e7}) comes from the derivation of the generalized Darcy's law, which is not restricted to only the instantaneous response of flow to head gradients. In fact, the exponential function inside the integral operator represents an appropriated kernel encompassing ballistic diffusion for short times and normal diffusion for long times \cite{HNV13}. These features are useful for modeling flow processes in highly heterogeneous porous media \cite{Dentz2006}.

For the second case, it is not possible to define a constant flow rate at the wellbore because of the non-locality induced by fractality and memory processes, so an asymptotic function is proposed which becomes constant as time increases. This is given by
\begin{equation}\label{e8}
Q(t) =1-\exp(-\frac{t}{\tau}).
\end{equation}
As a matter of fact, this condition is more realistic in practice than the one commonly used in drawdown well test analysis, which states a constant flow rate.

By selecting Eq. (\ref{e7}) for flow rate, one finds that the boundary condition must satisfy
\begin{equation}\label{e9}
\frac{\partial h}{\partial r}\big{|}_{r=1} = -1
\end{equation}
such that Eq. (\ref{e8}) is a solution of (\ref{e7}).
\section{Methodology} \label{method}
\subsection{Laplace transform}
The LT for any time-dependent function $h(r,t)$ is given by
\begin{equation}\label{e10}
\bar{h}= \mathcal{L}(h(r,t)) = \int_{0}^{\infty} h(r,t)e^{-st}ds,
\end{equation}
where $s$ is the LT variable. Some useful properties of LTs are: $\mathcal{L}(h_{t}(r,t)') = s\bar{h}-h(r,0)$ and $\mathcal{L}(h_{tt}(r,t)'') = s^{2}\bar{h}-sh(r,0)-h_{t}'(r,0)$, where $h_{t}(r,t)'$ and $h_{tt}(r,t)''$ are the first- and second-order partial derivatives with respect to time, respectively.

By applying the LT to Eq. (\ref{e1}) and using its respective initial conditions given by (\ref{e3}), the following transformed equation is obtained:
\begin{equation}\label{e11}
(1-\omega)\left(s\bar{h}_{1}-g_{1}(r)\right) = \frac{\lambda}{r^{d_{de}-d}}\left(\bar{h}_{2}-\bar{h}_{1}\right),
\end{equation}
and a relationship between $h_{1}$ and $h_{2}$ is derived as

\begin{equation}\label{e12}
\bar{h}_{1} = \frac{\bar{h}_{2}\lambda+g_{1}(r)(1-\omega) r^{d_{de}-d}}{s(1-\omega) r^{d_{de}-d}+\lambda}.
\end{equation}
On the other hand, applying the LT to Eq. (\ref{e2}) and using its respective initial conditions given by equations (\ref{e4}), one obtains the following equation:
\begin{equation}\label{e13}
\begin{split}
\omega\tau \left( s^{2}\bar{h}_{2}-g_{3}(r)-sg_{2}(r)\right)+\left(\omega + \lambda\tau r^{d-d_{bb}}\right)\left(s\bar{h}_{2}-g_{2}(r)\right)=\\
\frac{1}{r^{d_{bb}-1}}\frac{\partial }{\partial r}\left(r^{\beta}\frac{\partial\bar{h}_{2}}{\partial r}\right)+\left(\lambda\tau r^{d-d_{bb}}-(1-\omega)r^{d_{de}-d_{bb}}\right)\left(s\bar{h}_{1}-g_{1}(r)\right)
\end{split}
\end{equation}
where $g_{i}$, $i=1,2,3$, are the initial conditions for both dependent variables. After introducing (\ref{e12}) into (\ref{e13}) and some algebraic effort, considering that partial derivatives become ordinary derivatives in the Laplace domain, we obtain:
\begin{equation}\label{e14}
\frac{d}{dr}\left(r^\beta\frac{d\bar{h}_{2}}{dr}\right) -f_{s}(r)\bar{h}_{2}=\gamma_{s}(r)
\end{equation}
where function $f_{s}(r)$ is given by
\begin{equation}\label{e15}
\begin{split}
f_{s}(r) = \omega\tau s^{2}r^{d_{bb}-1}+\left(\omega r^{d_{bb}-1} + \lambda\tau r^{d-d_{bb}}\right)s-\\
\left(\lambda\tau r^{d-d_{bb}}-(1-\omega)r^{d_{de}-d_{bb}}\right)\frac{s\lambda}{s(1-\omega) r^{d_{de}-d}+\lambda},
\end{split}
\end{equation}
whereas $\gamma_{s}(r)$ is
\begin{equation}\label{e16}
\begin{split}
\gamma_{s}(r)= \left(\frac{s(1-\omega) r^{d_{de}-d}}{s(1-\omega) r^{d_{de}-d}+\lambda}-1\right)g_{1}(r)\left(\lambda\tau r^{d-d_{bb}}-(1-\omega)r^{d_{de}-d_{bb}}\right)+\\
 \omega\tau \left(g_{3}(r)+sg_{2}(r)\right)+ \left(\omega + \lambda\tau r^{d-d_{bb}}\right)g_{2}(r).
\end{split}
\end{equation}
Analytical solutions for Eq. (\ref{e14}) are only possible in some particular cases; for example, for $d_{bb}=d_{de}=d=2$ and for $\omega=1$, $\lambda=0$, and $d_{de}=2$. In the first case, we have a telegraphic Euclidean double-porosity model, while in the second we have a fractal telegraphic single-porosity model. Both have been solved elsewhere for an infinite reservoir \cite{HNV13}. Further, analytical solutions for infinite Euclidean non-telegraphic models, $\tau =0$, were studied in \cite{EH49,WR63,CFV08}. The semi-numerical solution presented here encompasses all mentioned cases that can be derived from Eqs. (\ref{e1}) and (\ref{e2}) through proper selection of parameters.

If the transformation $\xi=\int r^{-\beta}dr$ is proposed for the spatial coordinate, then the new space-independent variable must be evaluated based on the relation
\begin{equation}\label{e17}
\xi =
\begin{cases}
	\log(r) & \beta =1 \, , \\
	\frac{1}{1-\beta}r^{1-\beta} & \beta\, \#\, 1\, ,
\end{cases}
\end{equation}
so that Eq. (\ref{e14}) becomes
\begin{equation}\label{e18}
\frac{d^2\bar{h}_{2}}{d\xi^2} -\hat{f}_{s}(\xi)\bar{h}_{2}=\hat{\gamma}_{s}(\xi)
\end{equation}
where $\hat{f}_{s}(\xi)= b^{\beta/b}\xi^{\beta/b} f_{s}(\xi) $, $\hat{\gamma}_{s}(\xi)=b^{\beta/b}\xi^{\beta/b}\gamma_{s}(\xi)$ and $b= 1/(1-\beta$). In order to simplify algebraic manipulation, henceforth the subscripts for $\bar{h}_{2}$, $\hat{f}_{s}$, and $\hat{\gamma}_{s}$ are omitted.

\subsection{Finite differences in the Laplace domain}
Defining the approximation of the dependent variable as $\bar{h}_{i}\approx\bar{h}(\xi_{i})$ at the {\it i}th grid element, functions in (\ref{e18}) become $\hat{f}_{i}=\hat{f}(\xi_i)$ and $\hat{\gamma}_{i}=\hat{\gamma}(\xi_{i})$, and  by using centered finite differences to approximate the second-order derivative, i.e.,
\begin{equation}\label{e19}
\frac{d^2 \bar{h}}{d\xi^2}\approx \frac{\bar{h}_{i-1}-2\bar{h}_{i}+\bar{h}_{i+1}}{\Delta\xi^2},
\end{equation}
the discretized representation of (\ref{e18}) is
\begin{equation}\label{e20}
\bar{h}_{i-1}-(2+\Delta\xi^2 \hat{f}_{i})\bar{h}_{i}+\bar{h}_{i+1}=\Delta\xi^2\hat{\gamma}_{i}.
\end{equation}
The space domain is uniformly discretized as $\xi_{i} =\xi_{i-1}+\,d\xi$, $i=2,\dots,M$, $\xi_{1}=0$ for $\beta =1$ and $b$ otherwise. $d\xi = \log(r)/M$ for $\beta=0$ and $(L^{b}-r_{w}^{b})/bM$ otherwise. $M$ is the number of subdivisions for the $\xi$ coordinate. Note that Eq. (\ref{e20}) is only valid within the spatial domain for $2\le i \le M-1$. At the boundaries, $i=1$ and $i=M$, proper equations have to be used based on the case analyzed. For constant head at the wellbore, $ h=h_{w}$. If only the backbone is connected to the wellbore and the model solution is in terms of the backbone, then just one inner boundary condition is required. The LT of this inner boundary condition is $\bar{h}_{w} =  h_{w}/s$, which corresponds to the first spatial node, whose discretized equation is
\begin{equation}\label{e21}
\bar{h}_{1} = \frac{h_{w}}{s}.
\end{equation}

The corresponding LT for outer boundary condition, Eq. (\ref{e6}), is $\partial \bar{h}/\partial \xi = 0$, and the proper equation for the last node using backward finite difference schemes is
\begin{equation}\label{e22}
\bar{h}_{M-1} - \bar{h}_{M} = 0.
\end{equation}

The dimensionless instant fluid flow in the Laplace domain results from the application of LT to Eq. (\ref{e7})
\begin{equation}\label{e23}
\bar{Q} = -\frac{1}{s\tau+1}\,\frac{\partial\bar{h}}{\partial\xi}\big{|}_{\xi_{1}}.
\end{equation}
On the other hand, boundary conditions in the Laplace domain for asymptotically convergent flow are generated after applying the LT to Eqs. (\ref{e6}) and (\ref{e9}). As the outer boundary condition is similar to the former case, Eq. (\ref{e22}) can be used in the same manner for the last node of the spatial domain. The inner boundary condition can be derived from (\ref{e9}), and its discretized representation is
\begin{equation}\label{e24}
\bar{h}_{1}-\bar{h}_{2}=\frac{\Delta\xi}{s}.
\end{equation}

The hydraulic head at the wellbore is evaluated by Eq. (\ref{e23}) based on
\begin{equation}\label{e25}
\bar{h}_{1}=\Delta\xi\, (s\tau + 1)\bar{Q}+\bar{h}_{2},
\end{equation}
where $\bar{Q}$ is the constant asymptotic flow rate, which takes some predefined value as $t$ increases.
Note that Eqs. (\ref{e24}) and (\ref{e25}) are equivalent, and both of them can be used to represent the inner boundary conditions and hydraulic head evolution at the wellbore.

\subsection{Numerical inverse Laplace transform}
The inverse LT is evaluated numerically with Stehfest's algorithm \cite{S70}. The procedure consists of taking an approximation of the real part of the Laplace parameter\cite{MMH94,HO14}, $s$, in order to evaluate a specific time, $t$, as follows:
\begin{equation}\label{e26}
s_{\nu} = \frac{ln(2)}{t}\nu,\qquad \nu=1,\dots,N_{p}
\end{equation}
where $N_{p}$ is an even number of parameters used in the approach. For a particular time, the Laplace parameter is a vector, $\mathbf{s}$, of $N_{p}$ elements. Each element of $\mathbf{s}$, i.e., $s_{i}$, is substituted into discretized equations based on the particular case being solved. For constant head at the wellbore, $\mathbf{s}$ is used in Eqs. (\ref{e20})-(\ref{e22}), whereas, for asymptotically constant flow rate, $\mathbf{s}$ is substituted in Eqs. (\ref{e20}), (\ref{e22}), and (\ref{e24}).  The result is a tridiagonal linear algebraic system whose matrix form is 
\begin{equation}\label{e27}
\begin{bmatrix}
   {d_1} & {e_{1}} & {   } & {   } & { 0 } \\
   {1} & {d_2} & {1} & {   } & {   } \\
   {   } & {1} & {d_3} & \ddots & {   } \\
   {   } & {   } & \ddots & \ddots & {1}\\
   { 0 } & {   } & {   } & {c_{M}} & {d_M}\\
\end{bmatrix}_{\nu}
\begin{bmatrix}
   {h_1 }  \\
   {h_2 }  \\
   {h_3 }  \\
   \vdots   \\
   {h_M }  \\
\end{bmatrix}_{\nu}
=\Delta\xi^{2}
\begin{bmatrix}
   {\gamma_1 }  \\
   {\gamma_2 }  \\
   {\gamma_3 }  \\
   \vdots   \\
   {\gamma_M }  \\
\end{bmatrix}_{\nu}
\end{equation}
which, in compact notation, can be written as $ \mathbf{M}\,\mathbf{h} =\mathbf{\gamma}$, where $\mathbf{M}$ are the matrix coefficients, $\mathbf{h}$ is the unknown vector of the head-transformed $\bar{h}$, and $\mathbf{\gamma}$ is the known right-hand-side vector. Solution of the algebraic system is performed efficiently through Thomas' algorithm \cite{T49}, which is equivalent to
\begin{equation}\label{e28}
\mathbf{h}_{\nu} =\mathbf{M}^{-1}_{\nu}\mathbf{\gamma}_{\nu}\qquad \nu=1,\dots,N_{p}.
\end{equation}

The coefficients of $\mathbf{M}$ correspond to those in Eq. (\ref{e20}), where, for $2\le i\le M-1$, $d_{i}= -2-\Delta\xi^{2}\hat{f}_{i}$ and $c_{i}=e_{i}=1$. Coefficients for the first and last rows depend on the boundary conditions. For constant head at the wellbore and based on Eqs. (\ref{e21}) and (\ref{e22}), $d_{1}=1$, $e_{1}=0$, $\gamma_{1}=1/s_{\nu}+\Delta\xi^{2} \hat{\gamma}_{1}$, $c_{M}=1$, and $d_{M}=-1$. For asymptotically constant  flow rate, the only change is at the inner boundary as the other coefficients remain the same as in the former case. In this case $d_{1}=1$, $e_{1}=-1$,  and $\gamma_{1}=\Delta\xi/s_{\nu}+\Delta\xi^{2} \hat{\gamma}_{1}$. Note that $\hat{\gamma}_{1}$ depends on the initial conditions and,  based on the definitions of the dimensionless variables in \cite{HNV13}, they are zero; therefore, $\hat{\gamma}_{1}$ must be zero as well.\\
 
The solution arises from the summation of these individual solutions, $\mathbf{h}_{\nu}$, based on the following equations:
  \begin{equation}\label{e29}
 h(r,t)=\frac{ln(2)}{t}\sum_{\nu=1}^{Np}W_{\nu}\, \mathbf{h}_{\nu},
 \end{equation}
 \begin{equation}\label{e30}
 W_{\nu}=F_{\nu}\sum_{\kappa=L_{0}}^{L_{M}}\frac{\kappa^{\frac{N_{p}}{2}}(2\kappa)!}{\left(\frac{N_{p}}{2}-\kappa\right)! \kappa! (\kappa-1)!(\nu-\kappa)!(2\kappa-\nu)!},
 \end{equation}
 where $F_{\nu}=(-1)^{\frac{N_{p}}{2}+\nu}$, $L_{0}=(\nu+1)/2$, and $L_{M}=\mathtt{min}(\nu,N_{p}/2)$. Note that Eqs. (\ref{e29}) and (\ref{e30}) allow us to evaluate the flow rate given by (\ref{e23}) by changing $\mathbf{h}_{\nu}$ for $Q_{\nu}$. We must recall that the solution in (\ref{e29}) corresponds to the backbone hydraulic head $h_{2}$ in Eqs. (\ref{e1}) and (\ref{e2}), and that it is related to dead-ends through Eq. (\ref{e12}).
\subsection{Semi-analytic solutions}
In order to qualitatively validate the semi-numerical solutions, we define some model parameters in Eqs. (\ref{e1}) and (\ref{e2}) so that the semi-analytic solutions are feasible to obtain. Semi-analytic solutions are exact in space and approximate in time as a numerical inversion algorithm is needed to invert the solution from the Laplace domain. In this case, we set $\lambda=0$ and $\omega =1$, and the general semi-analytical solution in Laplace space is given by
\begin{equation}\label{e31}
	\bar{h}_{2}=r^{\frac{1-\beta}{2}}\left[A_{1}(s)I_{\upsilon}\left(\frac{\sqrt{\alpha}}{d_{w}}r^{d_{w}}\right)+A_{2}(s)K_{\upsilon}\left(\frac{\sqrt{\alpha}}{d_{w}}r^{d_{w}}\right)\right]
\end{equation}
where $ \alpha=\tau s^2 +s $, $ d_{w}=(\theta +2)/2 $ and $ \upsilon =(1-\beta)/2d_{w} $. $ I_{\upsilon}$ and $ K_{\upsilon} $ are the modified Bessel functions of order $ \upsilon $ of the first and second kinds, respectively. The {\it constants} $A_{1}(s)$ and $A_{2}$(s) in the solution depend on the boundary conditions and must be evaluated for each case. Note that this general solution encompasses different cases related to several parameter combinations, and even though we deactivated parameters connected with dual porosity for the fractal case, the solution is also valid for the telegraphic Euclidean dual-porosity system where $ d=d_{bb}=d_{de}=2 $ and $ \theta =0 $; then $d_{w}=1$ and $\upsilon =0$. Particular solutions are not presented in this work as they can be derived from Eq. (\ref{e31}) by using the appropriate boundary conditions.

\section{Results and discussion}\label{resdis}
In Table \ref{tab1}, we summarize several cases validated for two types of volumetric reservoirs whose analytic solutions are feasible to obtain in the Laplace domain: i) telegraphic Euclidean dual porosity and ii) telegraphic fractal single porosity. The model solutions are derived for two different inner boundary conditions: a) constant hydraulic head and  b) asymptotically constant flow rate. Meanwhile, the outer boundary condition obeys zero flux. For constant head at the wellbore, the flow rate is evaluated whereas, for asymptotically constant flow rate, the hydraulic head dynamics are presented at the same location. For the first type of reservoir, we varied relevant dual-porosity parameters as $\omega$ and $\lambda$, while the remaining parameters were fixed. For telegraphic fractal single-porosity reservoirs, we set $\lambda =0$ and $\omega = 1$, and the model parameters related to fractality, $\theta$ and $d_{bb}$, were varied. Based on the original definitions of the dimensionless variables in the general model, the initial conditions used in validation and study were all zero.

\begin{table}[h!]
\caption{\label{tab1} Cases for validation with semi-analytical solutions for i) telegraphic Euclidean dual-porosity reservoirs, and ii) telegraphic fractal single-porosity reservoirs.}
\begin{tabular}{lll}
\hline
\hline
\multicolumn{3}{c}{i. Telegraphic Euclidean Dual-Porosity Reservoir ($\tau\neq 0$, $d_{bb}=d_{de}=2$, $\theta =0$, $\lambda \neq 0$, $\omega\neq 0$)} \\
\hline
\hline
& $h_{2}$ & $Q$ \\ 
$\omega=\{ 0.1, 0.4, 0.8\}$, $\lambda = 10^{-3}$, and $\tau = 10^{0}$ & Fig. \ref{fig:1}a & Fig. \ref{fig:1}b \\ 
$\lambda =\{ 10^{-1}, 10^{-5}, 10^{-9} \} $,  $\tau = 10^{0} $, and $\omega = 10^{-1} $ & Fig. \ref{fig:1}c & Fig. \ref{fig:1}d \\ 
$\tau  = \{ 10^{0}, 10^{1}, 10^{2} \} $, $\lambda =10^{-3}$, and $\omega = 10^{-1} $ & Fig. \ref{fig:1}e & Fig. \ref{fig:1}f \\ 
\hline
\hline
\multicolumn{3}{c}{ii. Telegraphic Fractal Single-Porosity Reservoir ($\tau \neq 0$, $d_{bb} \neq 0$, $d_{de}=2$, $\theta \neq 0$, $\lambda = 0$, $\omega =1 $)} \\
\hline
\hline
$\theta=\{ 0.05, 0.1, 0.3\} $, $\lambda =0$, $\omega =1$, $\tau = 10^{1} $,  and $d_{bb} = 1.95$ & Fig. \ref{fig:2}a & Fig. \ref{fig:2}b \\ 
$d_{bb}=\{ 1.5, 1.6, 2.0\} $,  $\lambda =0$, $\omega = 1$, $\tau = 10^{1}$, and $\theta = 0.1$ & Fig. \ref{fig:2}c & Fig. \ref{fig:2}d\\
\hline
\end{tabular}
\end{table}

\begin{table}	
\caption{\label{tab2} Cases for the fractal telegraphic dual-porosity model whose analytical solution is not feasible.}	
	\begin{tabular}{lll}
		\hline
		\hline
		\multicolumn{3}{c}{Fractal telegraphic dual porosity ($\tau\neq 0$, $d_{bb}\neq 0$, $d_{de} \neq 0$, $\theta \neq 0$, $\lambda \neq 0$, $\omega \neq 1 $) } \\ 
		\hline
		\hline
		& $h_{2}$ & $Q$ \\ 
		$\theta = \{0.1, 0.3, 0.5 \}$, $d_{bb} = d_{de} =1.8$ $\omega = 0.5$, $\lambda = 10^{-6}$, and  $\tau = 10^{1}$ & Fig \ref{fig:3}a & Fig \ref{fig:3}b \\ 
		
		$d_{de} = \{1.6, 1.8, 2.0 \}$,  $d_{bb} = 1.8$, $\theta = 0.2$,  $\omega =0.5$,  $\lambda = 10^{-6}$, and  $\tau = 10^{1}$ & Fig \ref{fig:3}c & Fig \ref{fig:3}d \\
		
		$d_{bb} = \{1.6, 1.8, 2.0 \}$,  $d_{de} = 1.8$, $\theta = 0.2$,  $\omega =0.5$,  $\lambda = 10^{-6}$,  and  $\tau = 10^{1}$ & Fig \ref{fig:3}e & Fig \ref{fig:3}f\\ 
		\hline	
	\end{tabular}
\end{table}

In Table \ref{tab2}, we summarize solutions for the general case: the fractal telegraphic dual-porosity model whose solution, to the best of our knowledge, has not been reported previously. As for validation cases, some model parameters are set and others are modified so that their effects can be isolated. The parameters fixed are related to dual porosity and telegraphic effects ($\omega$, $\lambda$ and $\tau$, respectively). We then vary those linked to fractality, such as $d_{bb}$, $d_{de}$, and $\theta$. In the first column on the right side of Tables \ref{tab1} and \ref{tab2} we list the figures where comparison of semi-numerical and semi-analytical dimensionless head ($h_{2}$) and flow ($Q$) are plotted.

For semi-numerical solutions, we used $M=10^{4}$ nodes for spatial discretization and $N_{p}=12$ terms for the Laplace parameter in the Stehfest series during numerical inversion. The semi-numerical and semi-analytic solutions were implemented and tested in FORTRAN 90. Semi-analytic solutions were generated by numerically inverting Eq. (\ref{e31}) using the same number of parameters in the Stehfest series as in the semi-numerical case. In the following figures, the semi-analytical solutions are denoted by dashed red curves whereas semi-numerical solutions are shown as continuous blue lines.

\begin{figure}[h!]
	\centering
	\includegraphics[width=1.855in]{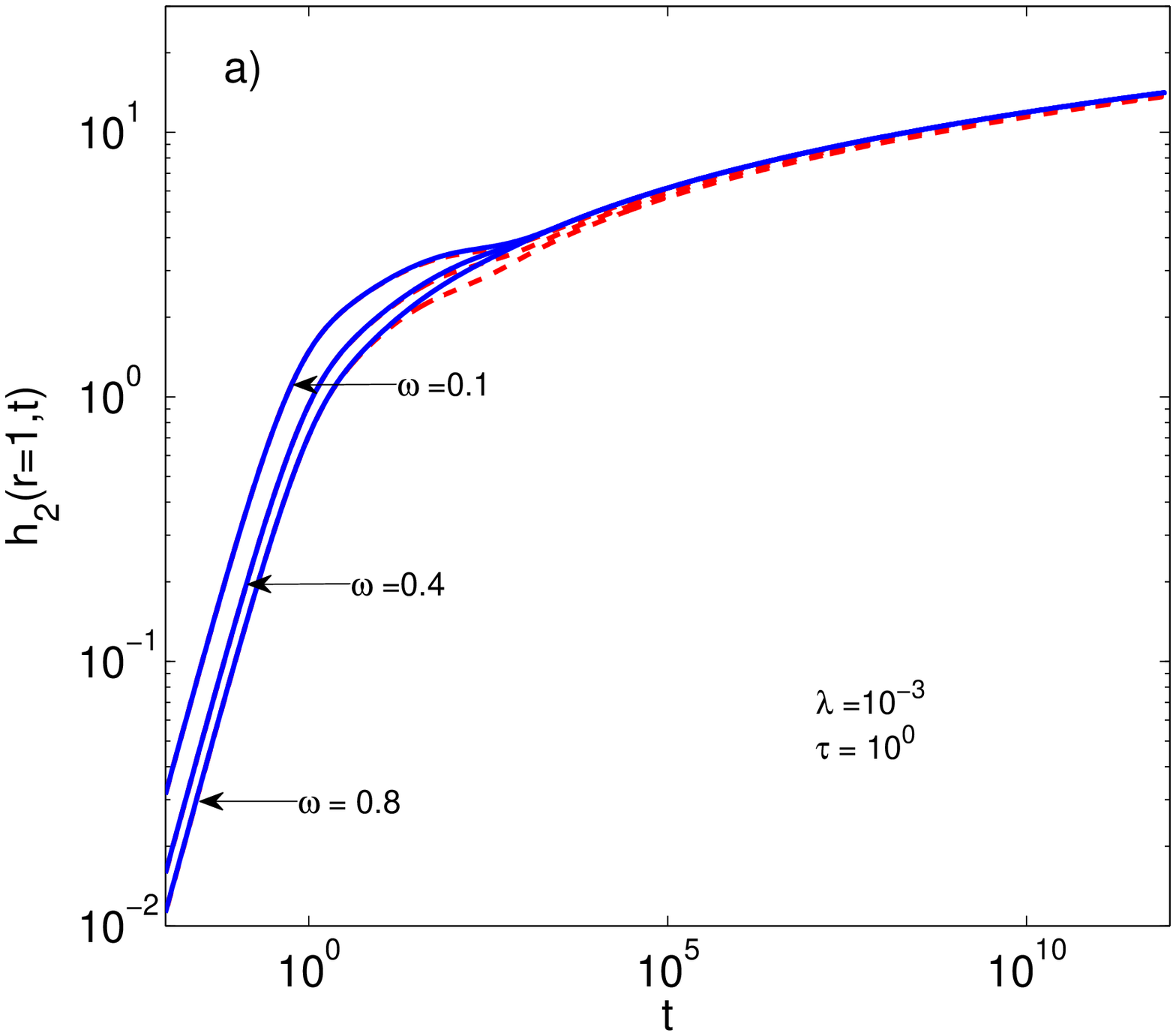}
	\includegraphics[width=1.855in]{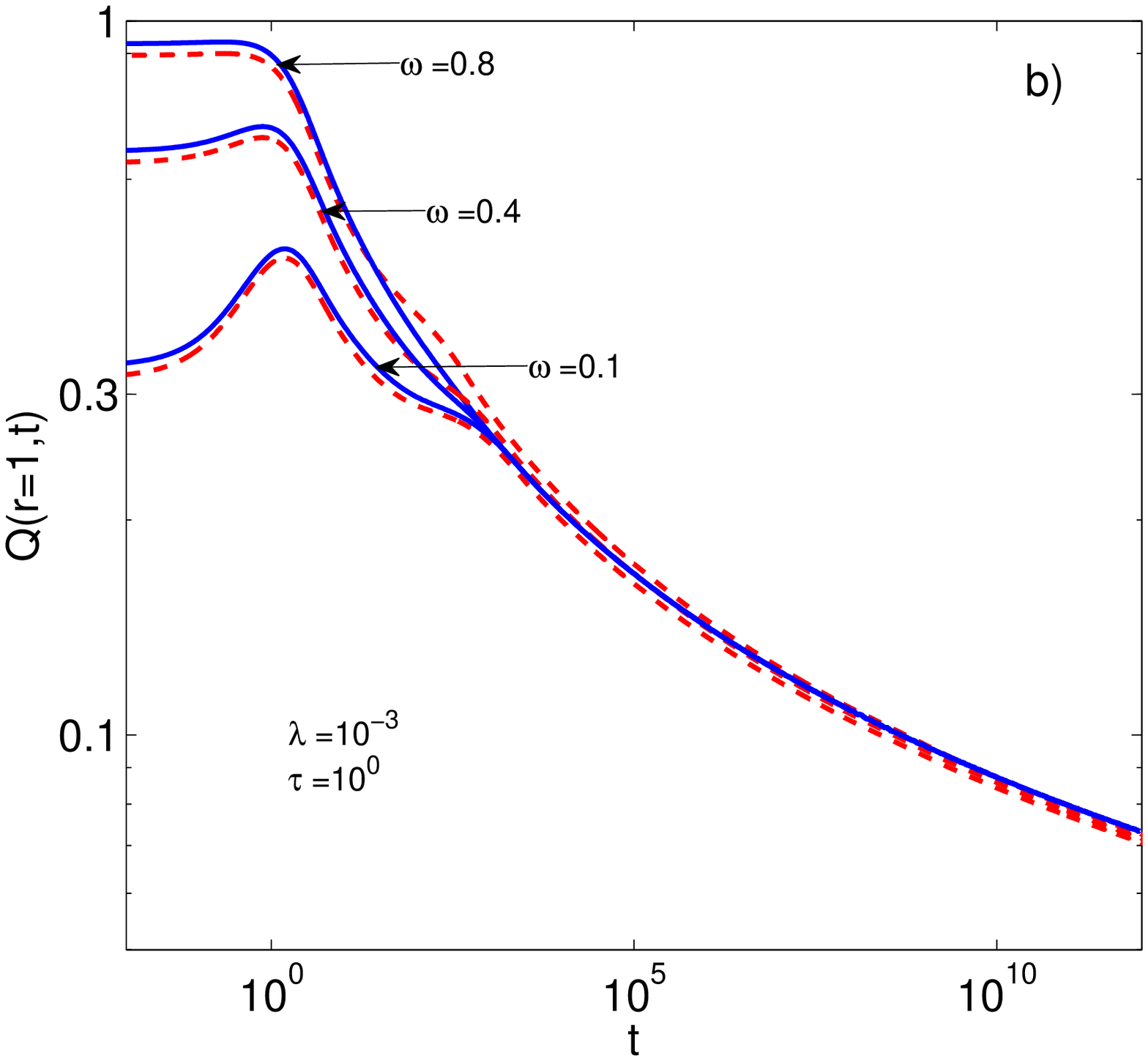}
	\includegraphics[width=1.855in]{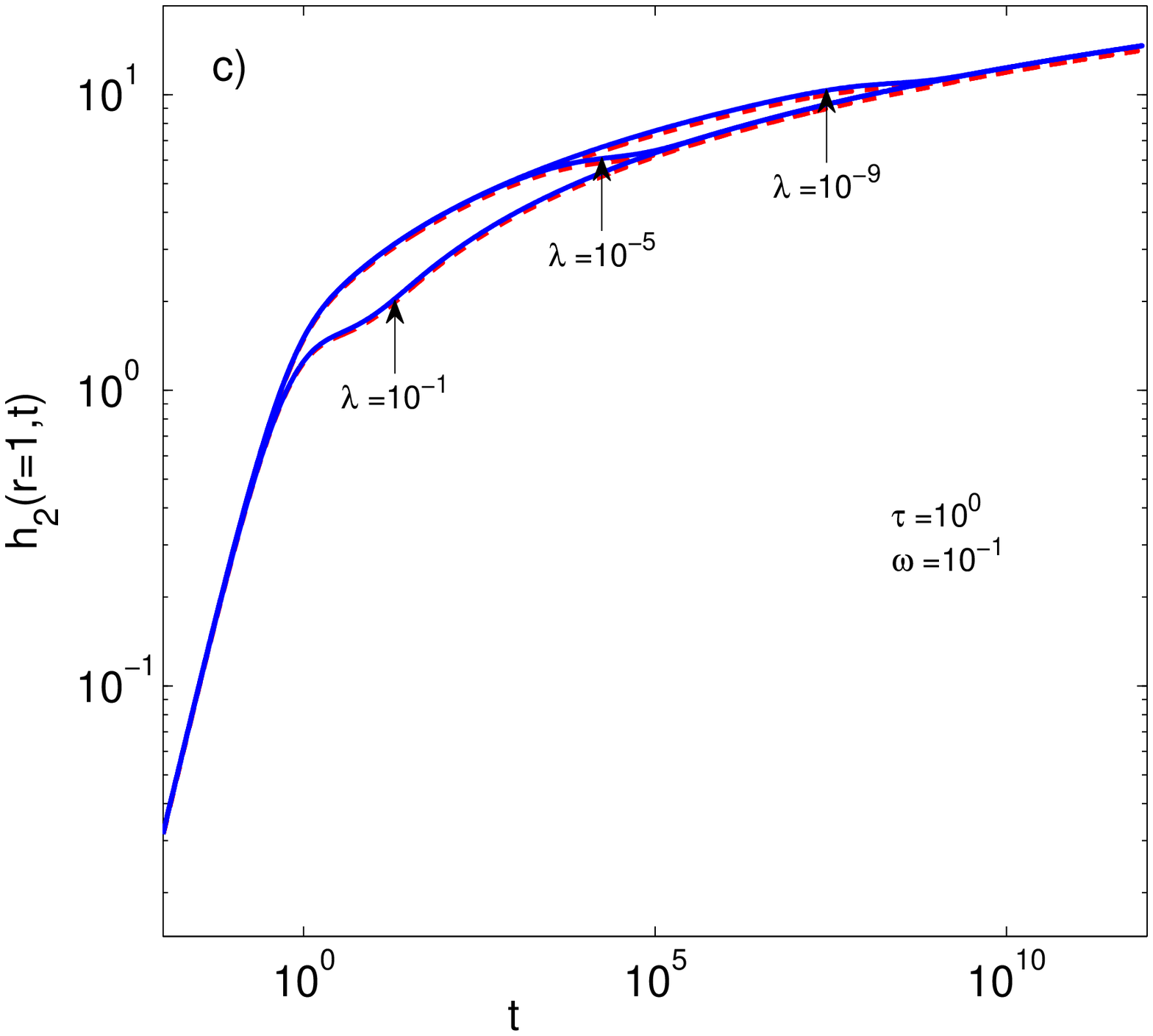}\\
	\includegraphics[width=1.855in]{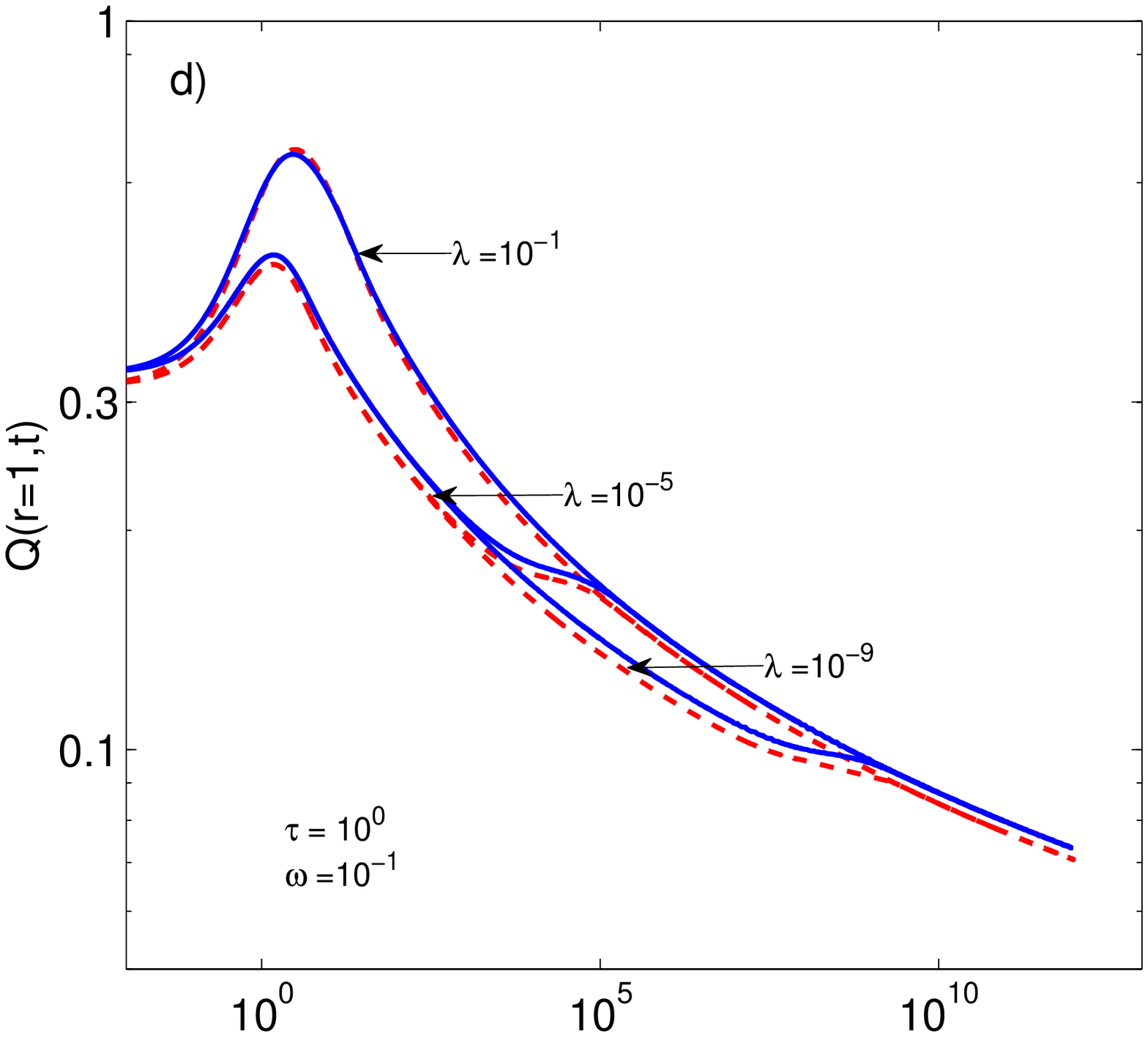}
	\includegraphics[width=1.855in]{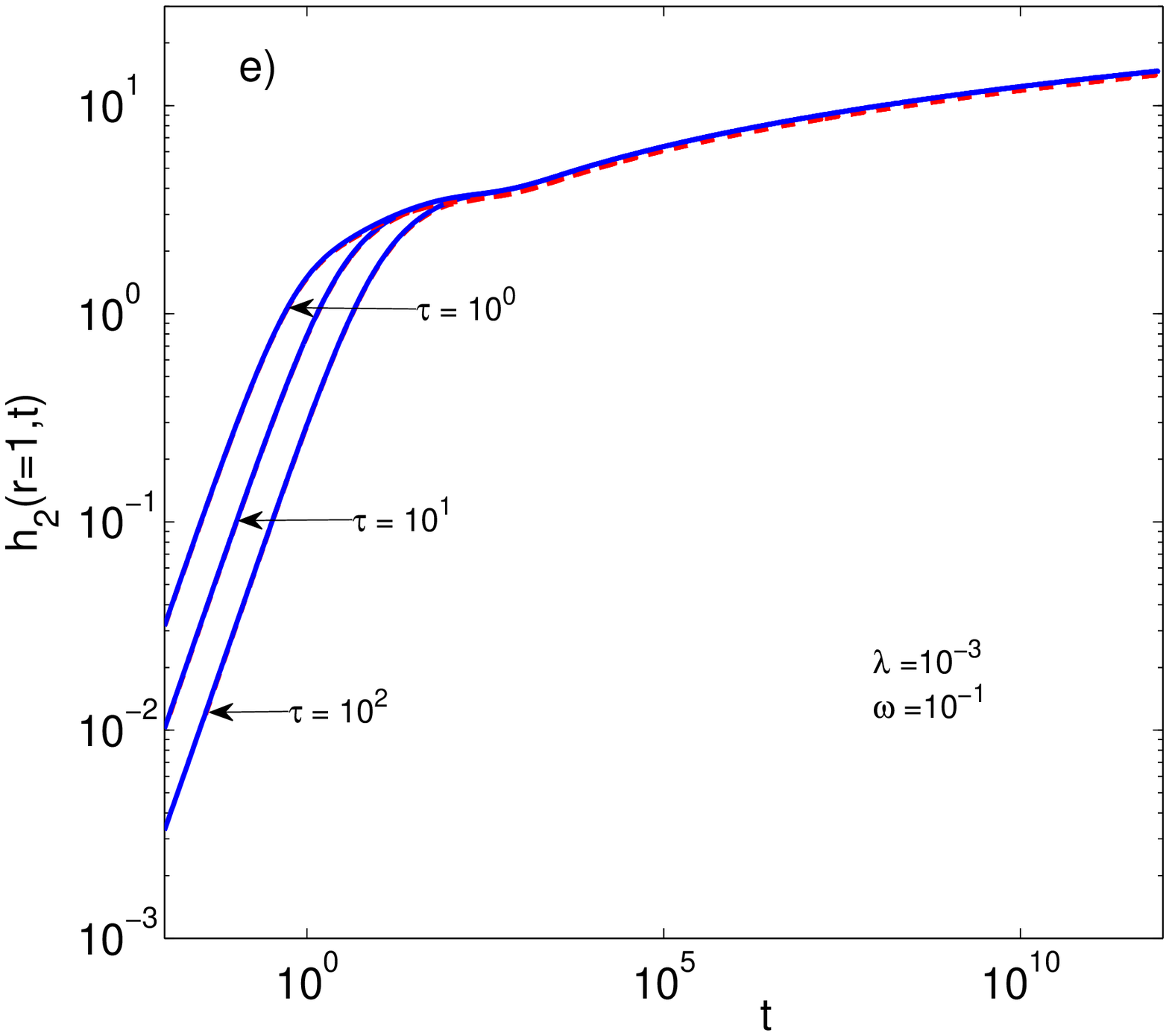}
	\includegraphics[width=1.855in]{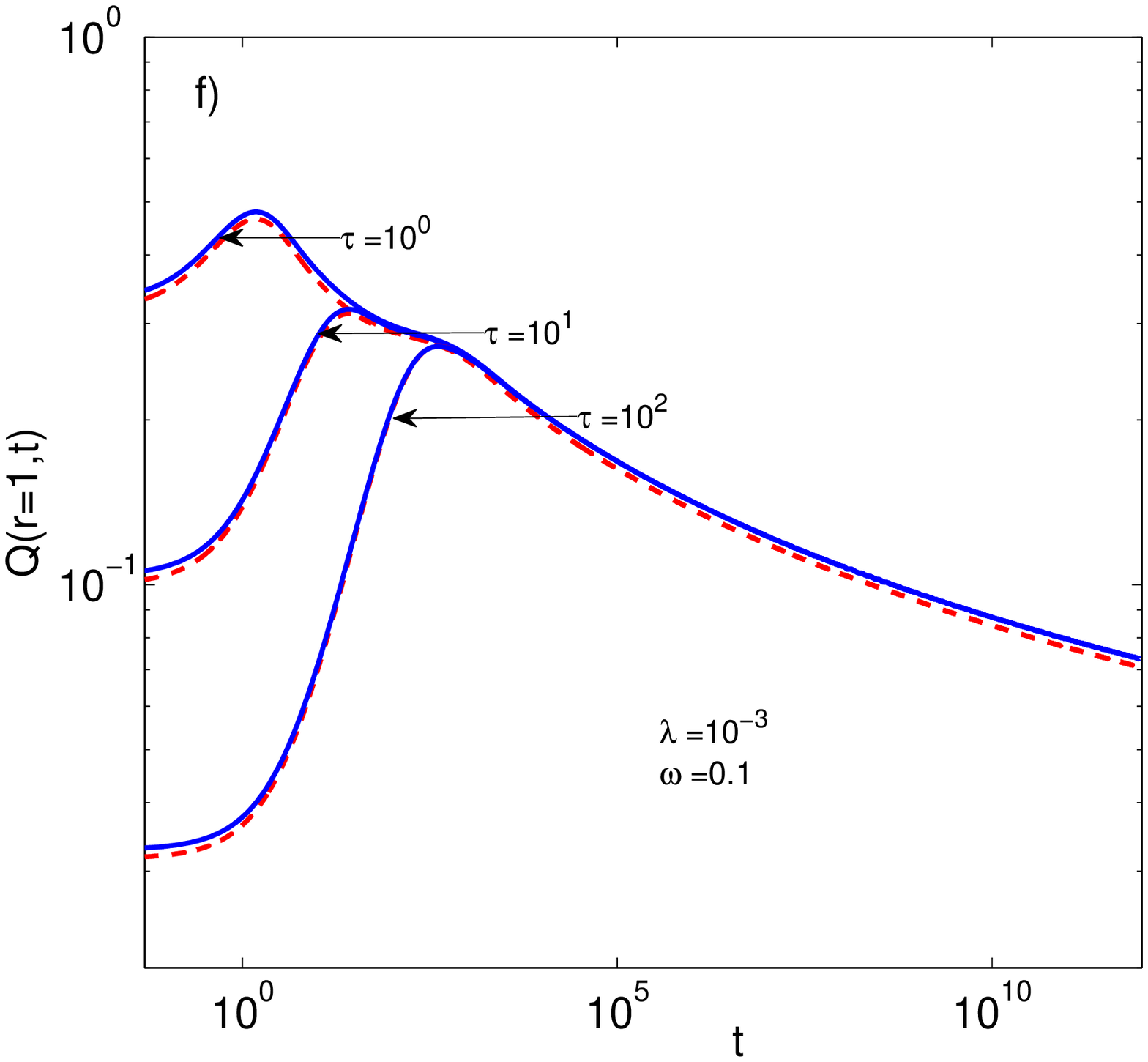}
	\caption{(Color online) Validation cases for a telegraphic Euclidean dual-porosity system ($d_{bb}=d_{de}=2$ and $\theta = 0$). Comparison between semi-numerical (continuous blue line) and semi-analytic solutions (dashed red line) for dimensionless head ( a), c), and e) ), and flow ( b), d) and f) ). The inner boundary at the wellbore was subjected to constant head and asymptotically constant flow. A zero-flux condition was imposed at the outer boundary.}
	\label{fig:1}
\end{figure}

In Fig. \ref{fig:1}, the effects of some model parameters over dimensionless head and flow are depicted. The semi-numerical case with semi-analytic solutions are compared for constant head at the wellbore in Figs. \ref{fig:1}b, \ref{fig:1}d, and \ref{fig:1}e, so the dimensionless flow is depicted; meanwhile, for asymptotically constant flow rate, Figs. \ref{fig:1}a, \ref{fig:1}c, and \ref{fig:1}d plot the dimensionless hydraulic head. It is worth emphasizing that the dimensionless head at the fractures increases with time owing to the way it was defined (see Eq. $11$ in \cite{HNV13}). The simulations carried out correspond to drawdown tests subject to constant production rate; furthermore, the dimensional head decreases with time. Qualitatively, the general behavior is well-described by the semi-numerical solution. Nonetheless, the head and flow are slightly overestimated in most cases. Such discrepancies are larger for flow, while the head differs only for long times. Observe that a maximum flow rate is reached for almost all the cases presented, which is significantly affected by the relaxation time $\tau$ and the storativity ratio $\omega$. In a general sense, the change of parameters has significant effects on the model solution, and such effects are well-captured by the semi-numerical procedure; therefore, the LTFD method can be reliably used for further applications. As mentioned by Hern\'andez and coworkers \cite{HNV13}, there is a time when the heterogeneities are smoothed, leading to pseudo-homogeneous reservoir behavior. This time can be elucidated when the governing equation for $h_2$, Eq. (\ref{e2}), is decoupled from the matrix head. This gives the following expression:
\begin{equation}\label{e32}
t_h = \frac{1-\omega}{\lambda} r^{d_{de}-d}
\end{equation}
which depends on physical and geometrical parameters and position. In our case, the head and flow are evaluated at the well, $r=1$; hence, homogeneous behavior takes place at  $t_h= \frac{1-\omega}{\lambda}$, which is the same conclusion as in \cite{HNV13}. $t_h$ can be easily observed as the inflection points for curves plotted in Fig. \ref{fig:1}. These are $t_h =$ 200, 600, and 900 for  Figs. \ref{fig:1}a and \ref{fig:1}b, $t_h =$ 9, $9\times10^4$, and $9\times10^8$ for Figs. \ref{fig:1}c and \ref{fig:1}d, and $t_h =$ 900 for Figs. \ref{fig:1}e and \ref{fig:1}f.

\begin{figure}[h!]
\centering
\includegraphics[width=2.5in]{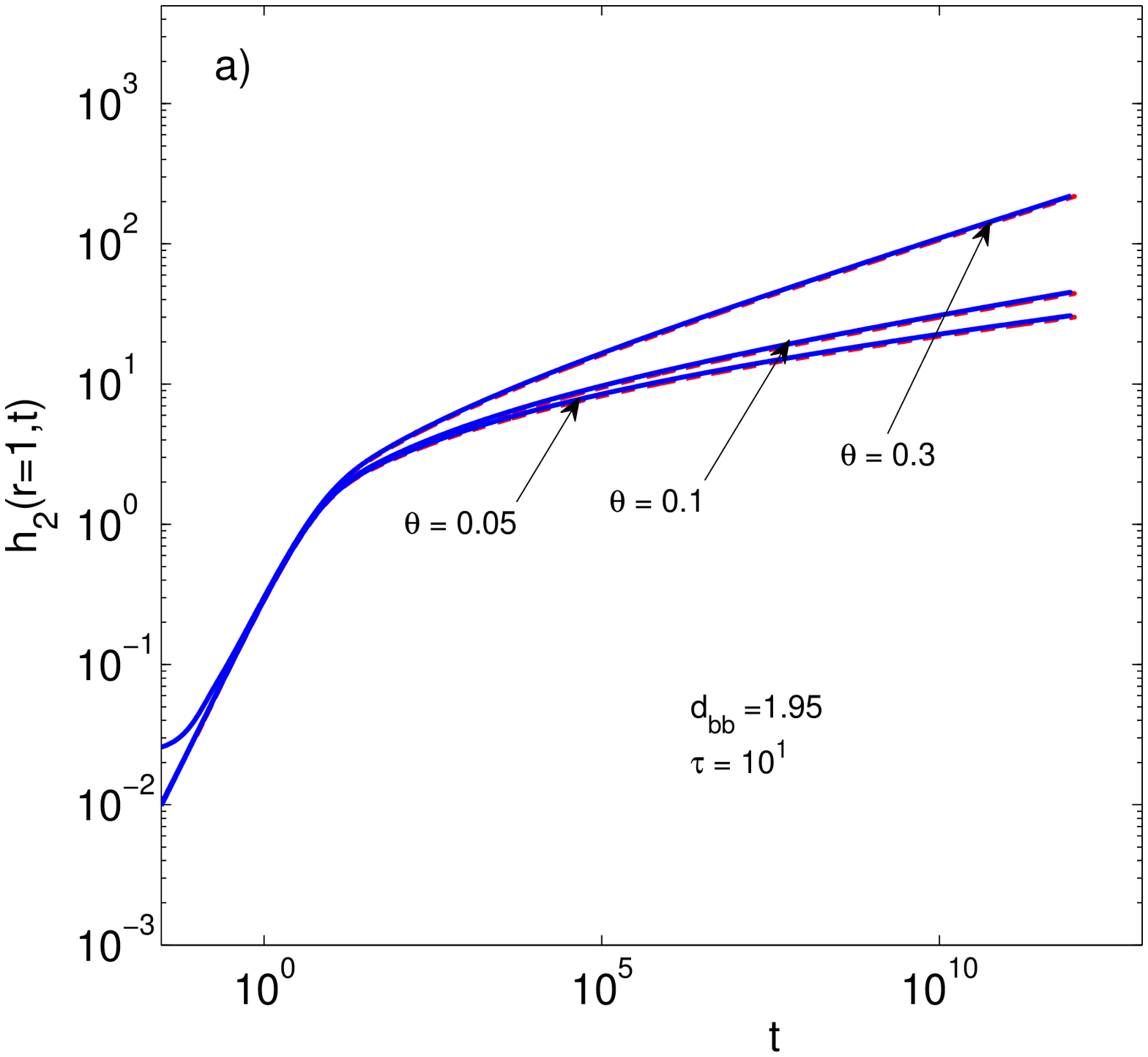}
\includegraphics[width=2.5in]{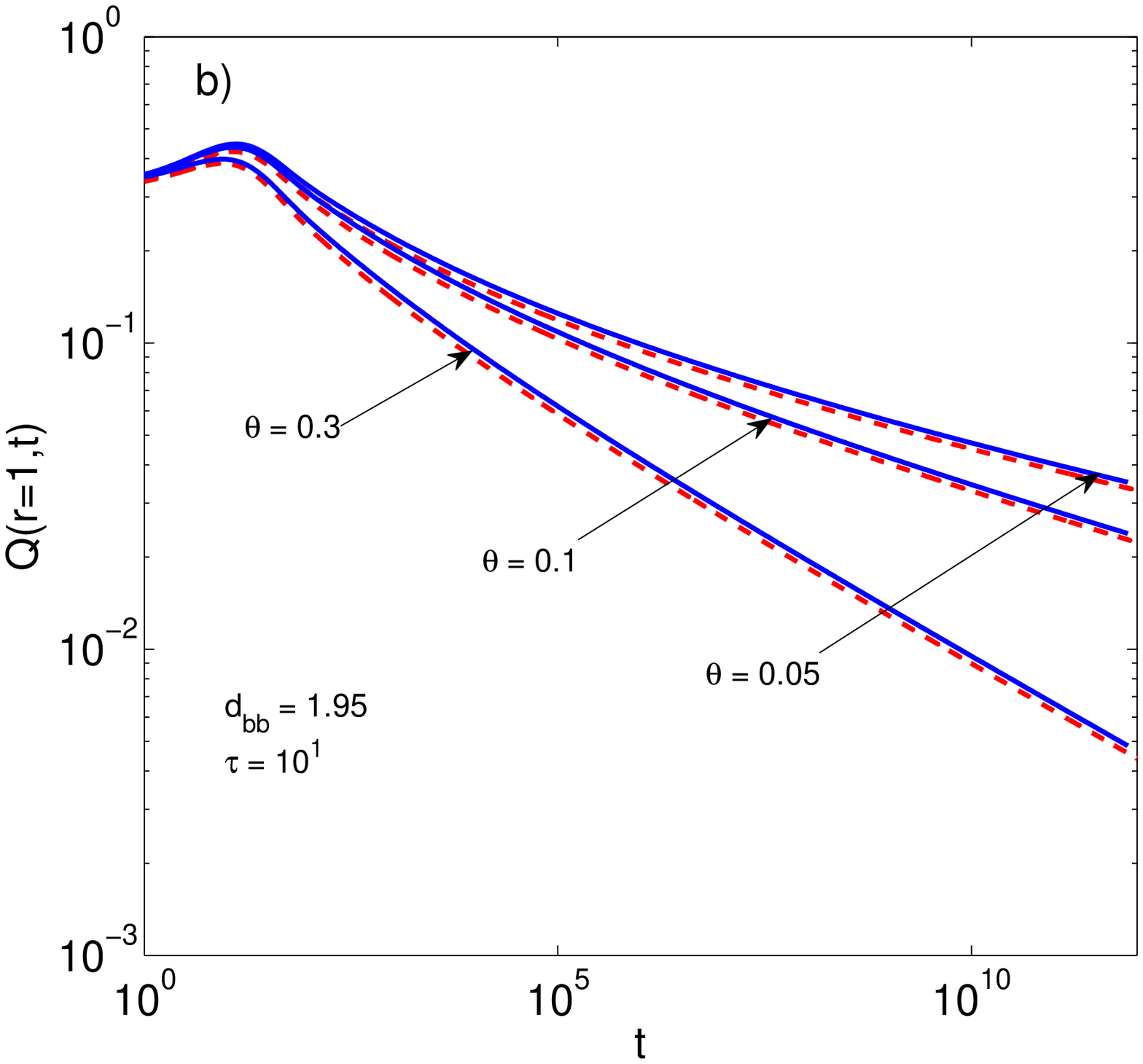}\\
\includegraphics[width=2.5in]{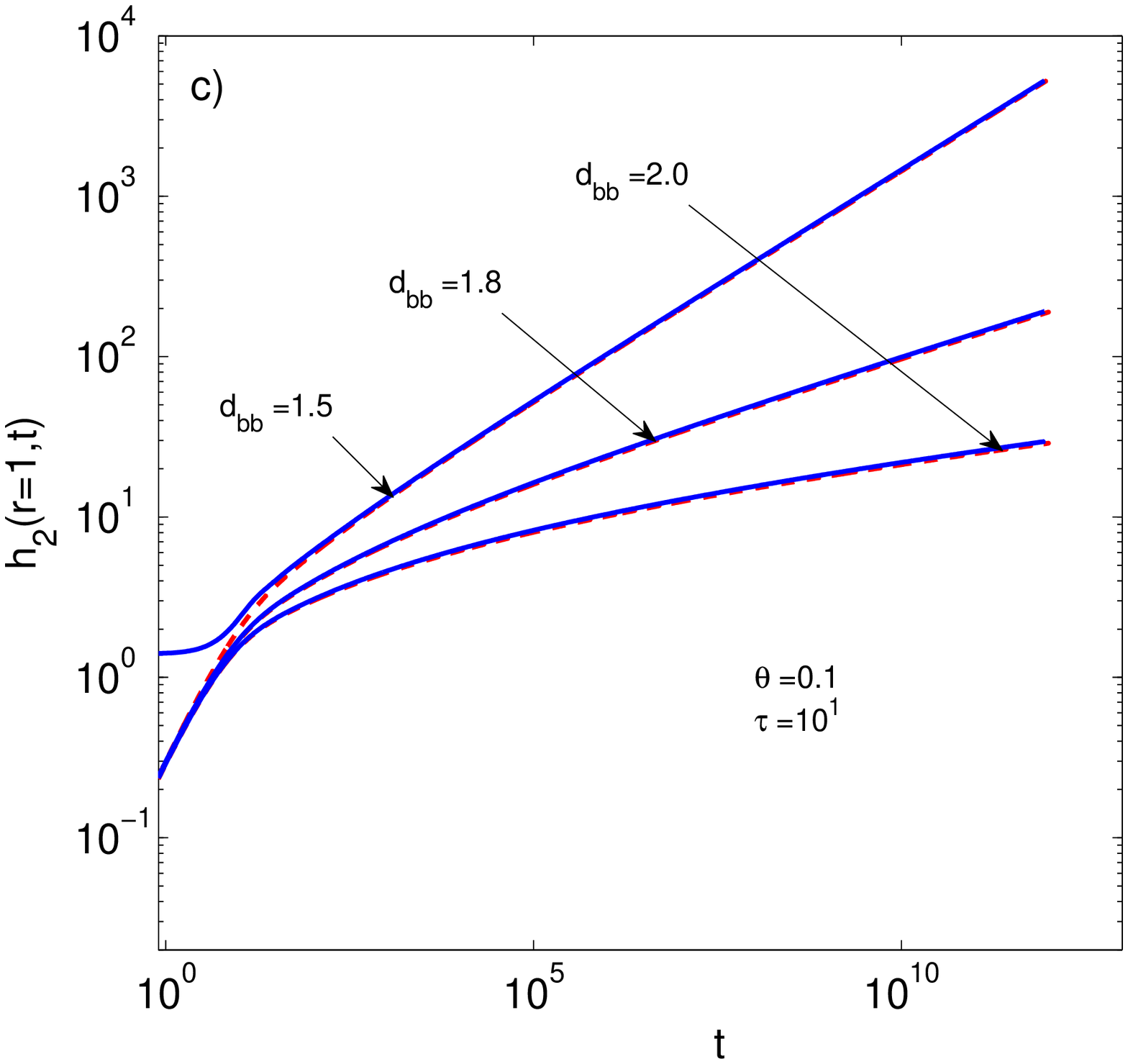}
\includegraphics[width=2.5in]{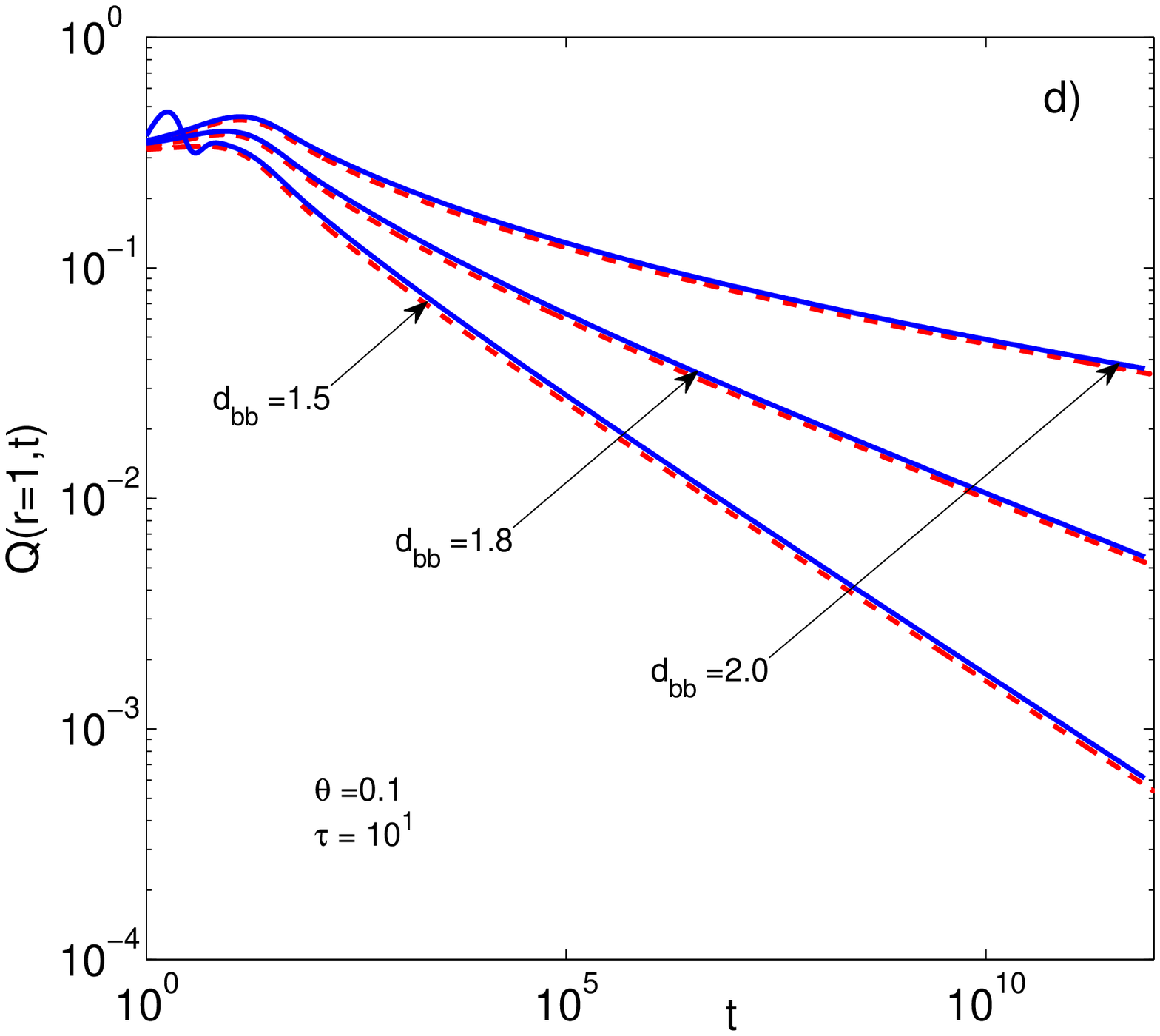}
\caption{(Color online) Validation cases for a telegraphic fractal single-porosity system ($d_{de}=2$, $\lambda =0$ and $\omega =1$). Comparison between semi-numerical (continuous blue line) and semi-analytic solutions (dashed red line) for dimensionless head ( a) and c) ), and for flow ( b) and f)). The inner boundary at the wellbore was subjected to constant head and asymptotically constant flow. A zero-flux condition was imposed at the outer boundary.}
\label{fig:2}
\end{figure}
The validated cases for telegraphic fractal single-porosity reservoirs are presented in Fig. \ref{fig:2}. The dimensionless head behavior for asymptotically constant flow at the wellbore is plotted in Figs. \ref{fig:2}a and \ref{fig:2}c, whereas the dimensionless flow for constant head at the same point is presented in Figs. \ref{fig:2}b and \ref{fig:2}d. The effects of the fractal model parameters, $\theta$ and $d_{bb}$, are exhibited therein. Once again, from a quantitative point of view, the semi-numerical and semi-analytical solutions match for almost all the cases analyzed. Nevertheless, small differences between solutions appear; these are mainly for flow, where the semi-numerical method slightly overestimates relative to semi-analytic results.  We must stress that short-term behavior seems more complicated to reproduce than long-term dynamic behavior. This can be explained by the manner in which time is treated in our methodology; as indicated in Eq. (\ref{e26}), a divergence appears in the Laplace domain as $t\to 0$. If the short-term behavior is of interest, an asymptotic solution may be derived from Eq. (\ref{e14}) by considering that $t\to 0$ implies $s\to \infty $; however, this is out of the scope of this work. Observe that unstable solutions are more evident when $d_{bb} = 1.5$, $\theta = 0.1$, and $\tau = 10$. Comparing results plotted in  Figs. \ref{fig:1} and \ref{fig:2}, one can conclude that the LTFD method is more suitable at any time for heterogeneous reservoirs where contributions of fluid from the matrix to the backbone network take place. For single-porosity reservoirs, one can exploit the capabilities of the LTFD method to model's long-time behavior for flow and head.

\begin{figure}[h!]
	\centering
	\includegraphics[width=2in]{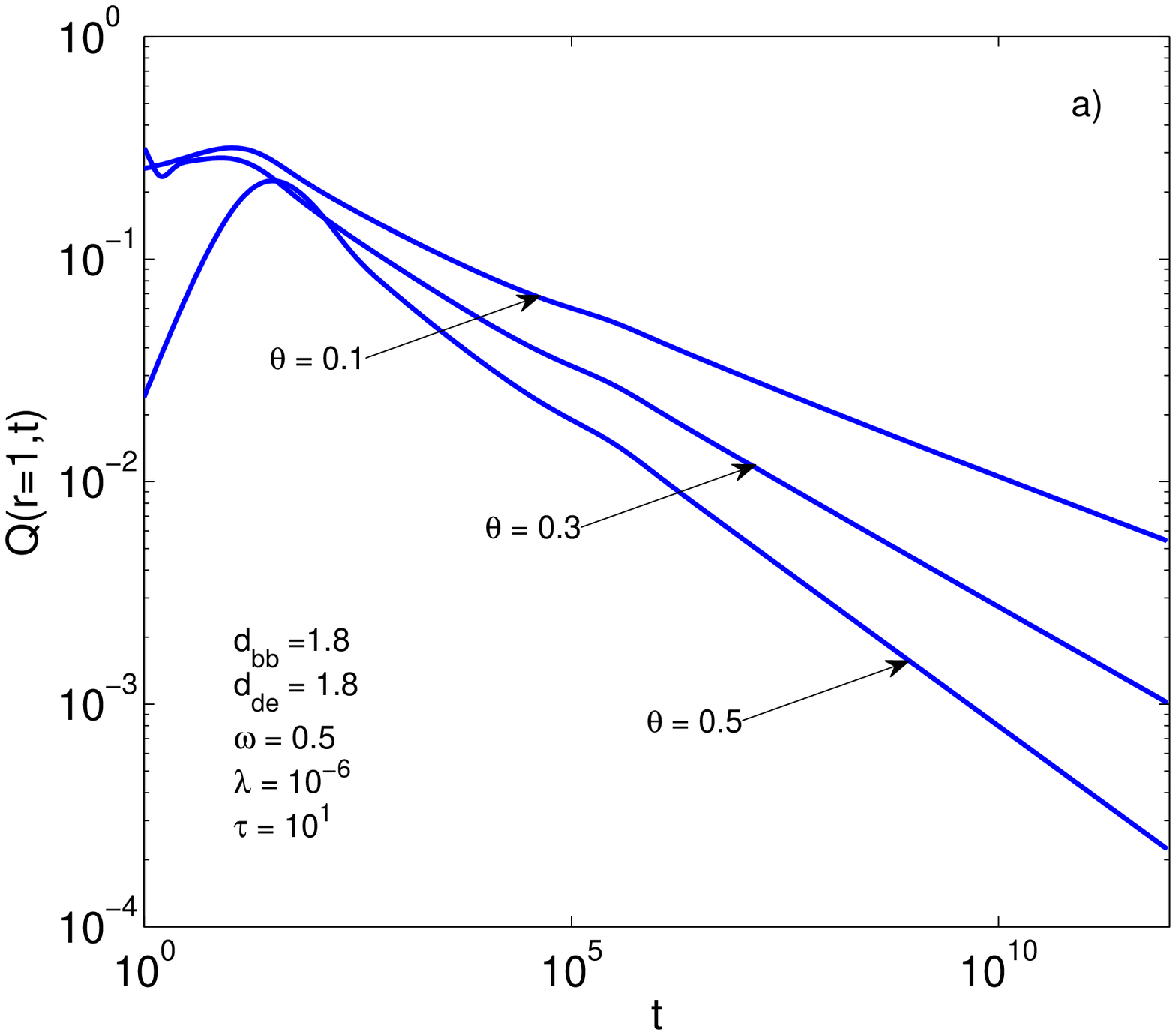}
	\includegraphics[width=2in]{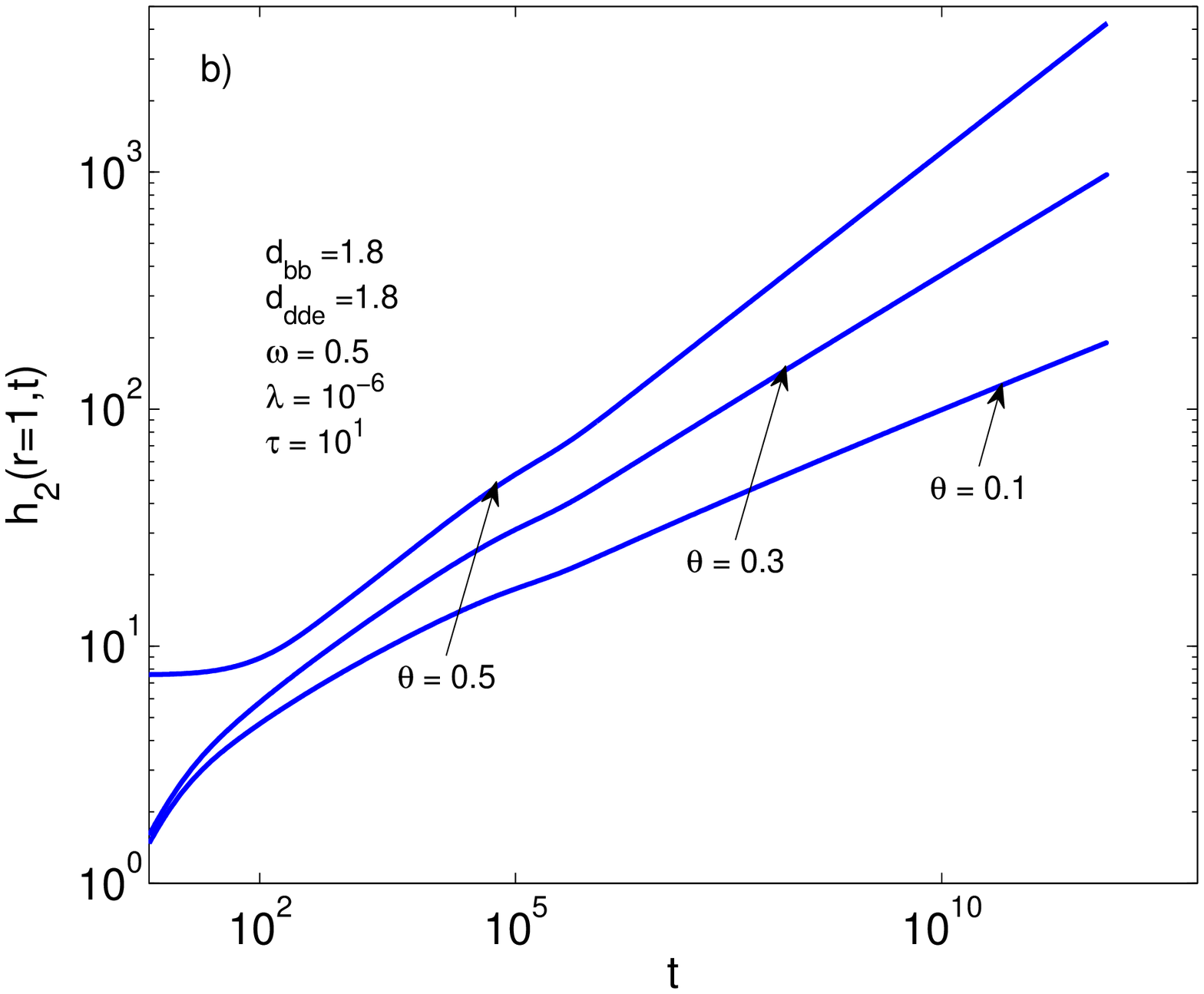}
	\includegraphics[width=2in]{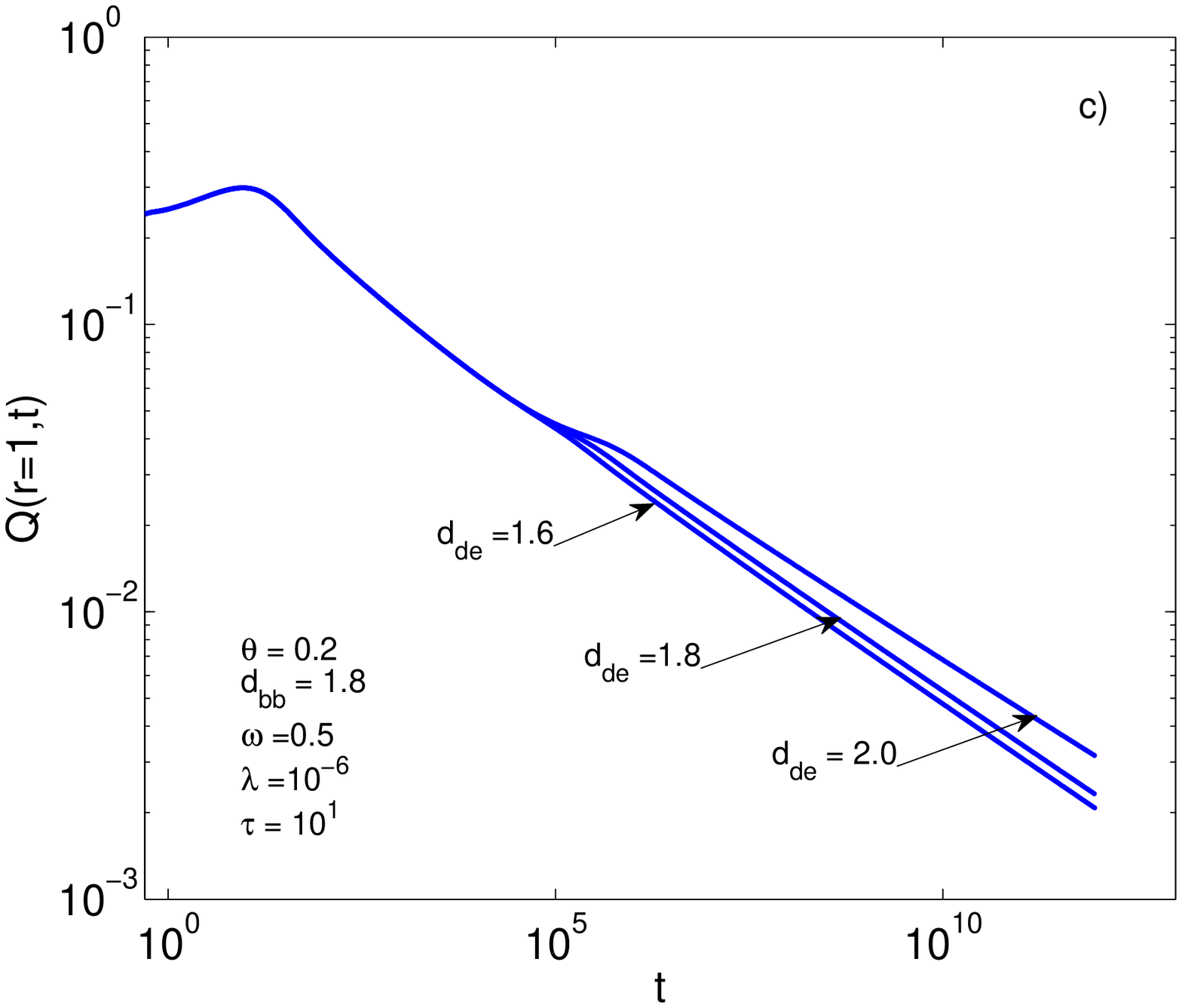}\\
	\includegraphics[width=2in]{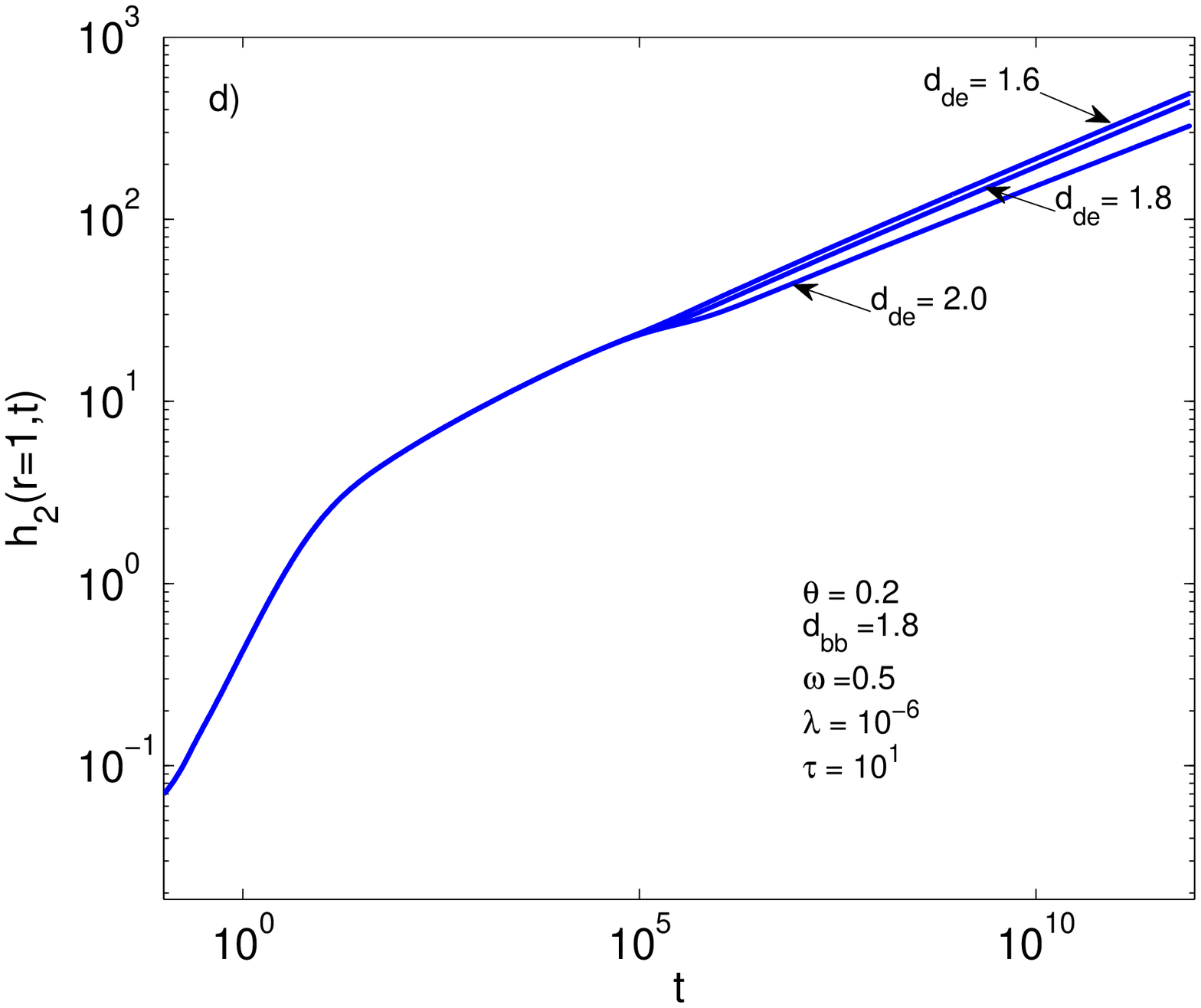}   
	\includegraphics[width=2in]{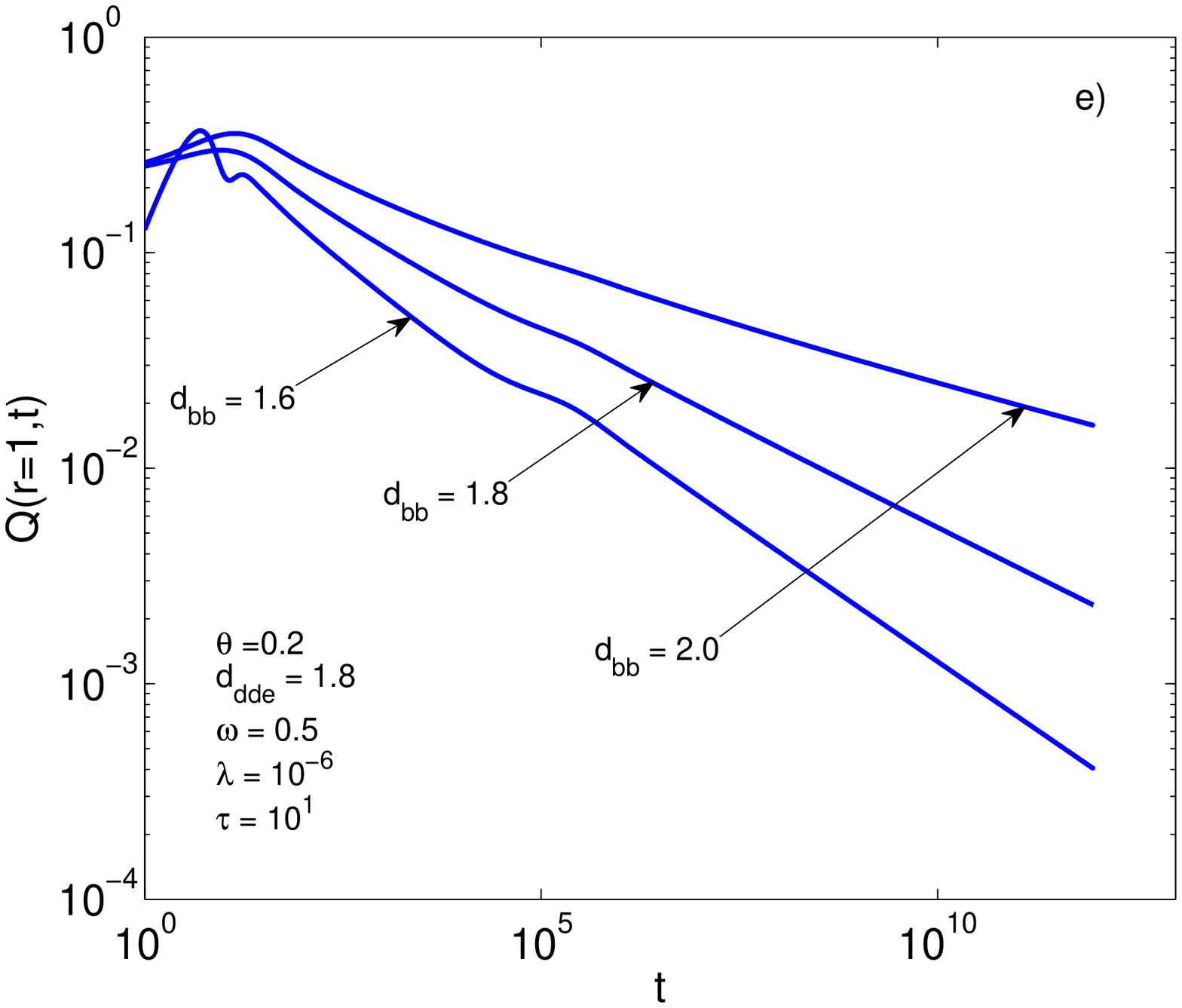}   
	\includegraphics[width=2in]{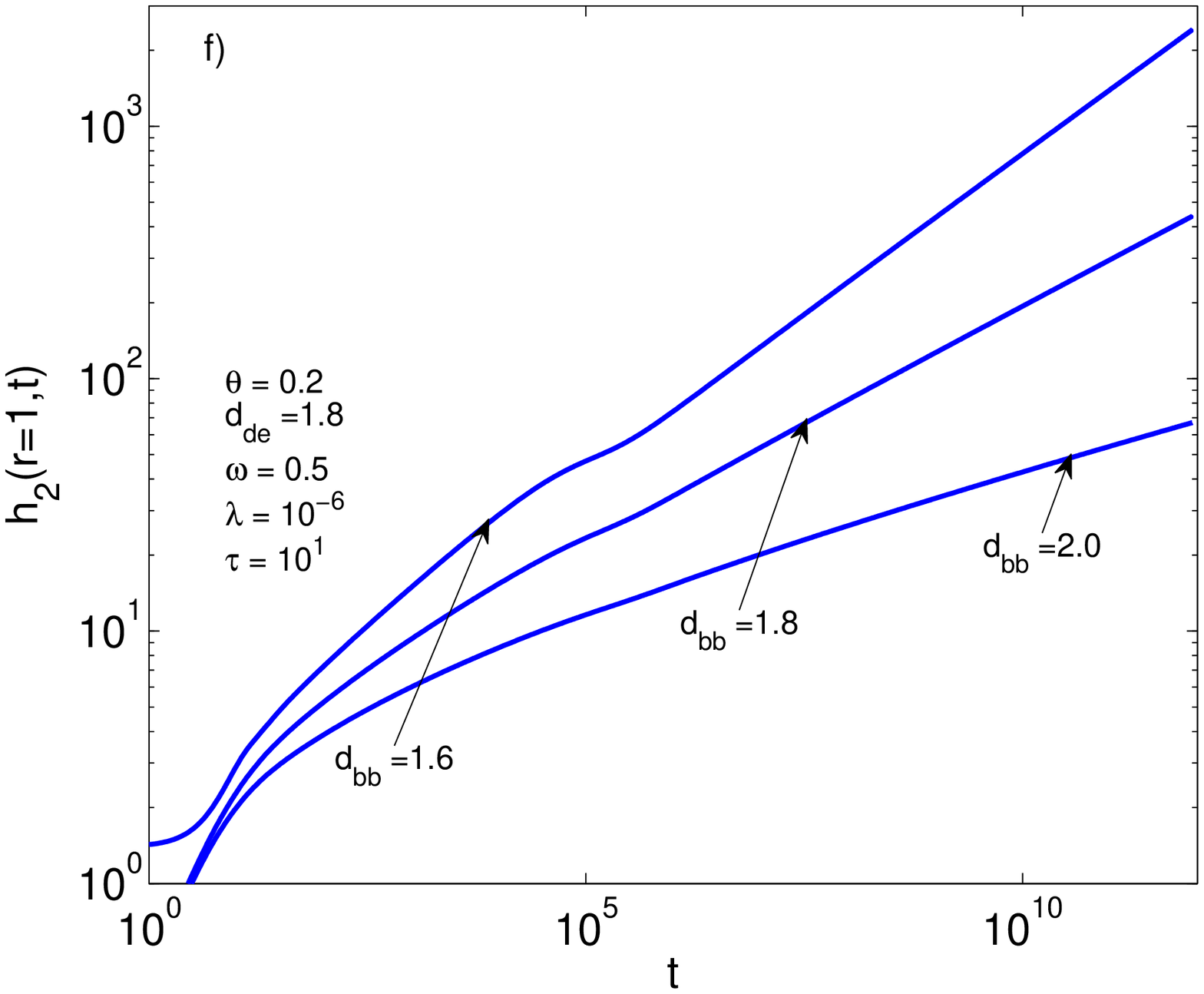}
		\caption{(Color online) Cases for a telegraphic fractal dual-porosity system. Here the dimensionless head at the wellbore is plotted in b), d), and f) for asymptotically constant flow rate at the inner boundary and zero flux at the external boundary. Meanwhile, the dimensionless flow for constant head at the wellbore and zero flux at the outer boundary is presented in a), c), and e).}
	\label{fig:3}
\end{figure}
Now we focus on the situation where solutions of the general model are not feasible either analytically or semi-analytically. Such cases were solved using the semi-numeric LTFD method, and some model solutions are plotted in Fig. \ref{fig:3}.  In our investigation, we varied the fractal parameters in order to explore their effects on flow performance and pay special attention to the physical meaning. Possible interconnections of fractal parameters were left aside as the main aim of this work is to present semi-numerical solutions. A deeper investigation of the dynamics of fractal non-Fickian reservoirs will be carried out in future, where physical connections between fractal parameters and rock fabric will be determined. Specially, the LTFD method will be applied to characterize aquifers, geothermal, and oil reservoirs constituted of fractured rock.

Results plotted in Fig. \ref{fig:3} exhibit characteristic regimes of dual-porosity models: short-, middle-, and long-term behaviors. As demonstrated above, for short times, small oscillations are observed in Figs. \ref{fig:3}a, b, e, and f, especially when the backbone fractal dimension and the permeability exponent differ from the Euclidean values, $\theta =0$ and $d_{bb} =2$, respectively.
As a matter of fact, the parameters $d_{bb}$ and $\theta$ have dominant effects over the head and flow in comparison with the dead-end fractal dimension $d_{de}$. An increase of $ \theta$ has the same effect as an equivalent decrease of $d_{bb}$. This can be inferred from comparison between Figs. \ref{fig:3}$a$ and \ref{fig:3}$b$, $f$ and $e$.
Results plotted in Figs. \ref{fig:3}$a$ and $b$ correspond to the flow regime lying between linear and radial dimensions (see the values of $d_{bb}$ and $d_{de}$) and for moderate backbone-dead-end interactions (see the values of $\omega$ and $\lambda$). At this point, it is worth recalling that $\theta$ quantifies effective connections in the backbone fracture network, where larger values indicate poor connectivity leading to smaller flow rate, as depicted in Fig. \ref{fig:3}$a$. This effect is also verified for substantial increments of the dimensionless head in Fig. \ref{fig:3}$b$, suggesting that it decreases significantly. Meanwhile, results plotted in Figs. \ref{fig:3}$c$ and $d$ indicate that fluid production is enhanced from the onset of homogeneous behavior, $t_h$, when $d_{de}$ increases. This effect is more noticeable in the case of the backbone fractal dimension. This phenomenon is explained, in part, by the fact that, in our simulations, there is little fluid exchanged between the backbone and dead-end networks; therefore, the main contribution to the flow rate comes from the backbone network. Summarizing, it is concluded that the closer the fractal dimension of dead-end and backbone structures are to the Euclidean dimension, the better the rate improvement obtained, while the opposite holds when the connectivity index, $\theta$, approaches unity.
\section{Conclusions}\label{conc}
In this paper, we have implemented a semi-numerical method to solve a fractal telegraphic dual-porosity fluid flow model whose coupled solution is not feasible to obtain semi-analytically. The transient part was handled using the Laplace transform technique, while spatial derivatives were discretized using central finite differences. Thus, step time constrictions are avoided as no discretization is required, and the space discretization fulfills the mass conservation without compromising numerical stability. As inversion from the Laplace domain to dimensionless time is not possible analytically, the Stehfest algorithm was used to compute the time evolution of flow rate and hydraulic head. Our approach was validated for analytically solvable cases, and good agreement was observed. It was found that larger values of fractal dimensions of dead-end and backbone networks improve the flow rate while the connectivity index diminishes the flow rate as it approaches unity. The capability of the methodology presented here encourages its application to characterizing fractured water, geothermal, and petroleum reservoirs.

\section*{Acknowledgments}
ECHH and DPL acknowledge support from CONACYT through C\'atedra at CIDESI.  ECHH and CGAM appreciate the funds granted by SEP-M\'exico through the research projects DSA/103.5/15/11043 and DSA/103.5/15/6797.

\end{document}